\documentclass[sigconf]{acmart} 

\AtBeginDocument{%
  \providecommand\BibTeX{{%
    \normalfont B\kern-0.5em{\scshape i\kern-0.25em b}\kern-0.8em\TeX}}}

\usepackage{verbatim}
\usepackage[normalem]{ulem}

\definecolor{amethyst}{rgb}{0.6, 0.4, 0.8}

\usepackage{soul}

\begin{document}

\title{Swinging in the States:\\Does disinformation on Twitter mirror\\ the US presidential election system?}


\author{Manuel Pratelli}
\email{manuel.pratelli@imtlucca.it}
\orcid{0000-0002-9978-791X}
\affiliation{%
  \institution{IMT Scuola Alti Studi Lucca}
  \streetaddress{Piazza San Francesco 19}
  \city{Lucca}
  \country{Italy}
  \postcode{55100}
}
\affiliation{%
  \institution{Istituto di Informatica e Telematica CNR}
  \streetaddress{via G. Moruzzi 1}
  \city{Pisa}
  \country{Italy}
  \postcode{56124}
}

\author{Marinella Petrocchi}
\email{marinella.petrocchi@iit.cnr.it}
\orcid{0000-0003-0591-877X}
\affiliation{%
  \institution{Istituto di Informatica e Telematica CNR}
  \streetaddress{via G. Moruzzi 1}
  \city{Pisa}
  \country{Italy}
  \postcode{56124}
}
\affiliation{%
  \institution{IMT Scuola Alti Studi Lucca}
  \streetaddress{Piazza San Francesco 19}
  \city{Lucca}
  \country{Italy}
  \postcode{55100}
}

\author{Fabio Saracco}
\email{fabio.saracco@cref.it}
\orcid{0000-0003-0812-5927}
\affiliation{%
  \institution{`Enrico Fermi' Research Center}
  \streetaddress{via Panisperna 89A}
  \city{Rome}
  \country{Italy}
  \postcode{00184}
}
\affiliation{%
  \institution{IMT Scuola Alti Studi Lucca}
  \streetaddress{Piazza San Francesco 19}
  \city{Lucca}
  \country{Italy}
  \postcode{55100}
}

\author{Rocco De Nicola}
\email{rocco.denicola@imtlucca.it}
\orcid{0000-0003-4691-7570}
\affiliation{%
  \institution{IMT Scuola Alti Studi Lucca}
  \streetaddress{Piazza San Francesco 19}
  \city{Lucca}
  \country{Italy}
  \postcode{55100}
}


\begin{abstract}
  For more than a decade scholars have been investigating the disinformation flow on social media contextually to societal events, like, e.g., elections.
  In this paper, we analyze the Twitter traffic related to the US 2020 pre-election  debate and ask whether it mirrors the electoral system.
  The U.S. electoral system provides that, regardless of the actual vote gap, the premier candidate who received more votes in one state `takes' that state. Criticisms of this system have pointed out that election campaigns can be more intense in particular key states to achieve victory, so-called {\it swing states}.
  Our intuition is that election debate may cause more traffic on Twitter-and probably be more plagued by misinformation-when associated with swing states. The results mostly confirm the intuition. 
About 88\% of the entire traffic can be associated with swing states, and links to non-trustworthy news are shared  far more in swing-related traffic than the same type of news in safe-related traffic.
  Considering traffic origin instead, non-trustworthy tweets generated by automated accounts, so-called social bots, are mostly associated with swing states. Our work sheds light on the role an electoral system plays in the evolution of online debates, with, in the spotlight, disinformation and social bots.
\end{abstract}



\maketitle

\section{Introduction}

In human society, disinformation and propaganda have a history as long as mankind. Such phenomena  manifest themselves most bitterly in times of crisis, such as wars and natural disasters\footnote{The long history of disinformation during war. Online: \url{https://www.washingtonpost.com/outlook/2022/04/28/long-history-misinformation-during-war/}. All urls have been accessed on October 13, 2022.}.
Consider, for example, the Battle of Actium in 31 B.C., one of the most momentous naval battles in history, which saw Marc Antony succumb and paved the way for Octavian to become the first emperor of the Roman empire, under the name Caesar Augustus. 
`Propaganda did not decide the war, but it allowed each side to rally its troops'~\cite{Strauss22war}. 


But while deception for societal strategies 
has always existed, the advent of the internet and social media has made the spread of false and biased news faster (we would say almost instantaneous), widespread, and `able to reach specific segments of the population'~\cite{Bradshaw2018junk}.  
In addition, the continuous bombardment of information from every medium to which we are subjected makes it extremely difficult to learn essential and correct information about an issue; so much so that the World Health Organization created a new term, {\it infodemic}, portmanteau  of the words {\it information} and {\it epidemic}, to define the information confusion to which we are exposed daily~\cite{cinelli2020info}.


Many scholars have focused on the 2016 U.S. presidential election, wondering how much digital disinformation influenced Trump's victory, but not finding a definitive answer. The study in~\cite{Georgacopoulos2020how} reports how, in the 3 months leading up to the election, fake news supporting Trump was shared on Facebook almost 4 times as many as the 8 million fake news supporting Clinton. Collecting more than 170 million tweets exchanged on Twitter in the 5 months prior to the same election, Bovet and Makse showed that the vast majority of news from trustworthy sources came from journalistic sources with verified Twitter accounts, while conspiracy theories, fake news and extremely biased news were mostly posted through unofficial Twitter clients, by unknown users, who then disappeared from Twitter, or automated accounts, {\it also known as} social bots~\cite{Bovet2019influence}. 
Shao et al., in~\cite{shao2018anatomy}, have highlighted the role of Twitter bots too, showing how bots were primarily responsible for the early spread of disinformation, interacting with influential accounts through mentions and replies.

Bots' evolutionary ability, namely the tendency they have acquired over the years to evade detection techniques,  is well known~\cite{Ferrara2016rise,Cresci2017paradigm}. The study in~\cite{luceri2019evolution}, for example, showed that, in the transition between the 2016 presidential elections to the 2018 midterms, the bots involved in the online political debate have evolved to the point where they are much more easily mistaken for humans.
Analyses have also been done on the manipulation of online narratives about the 2020 U.S. elections. Ferrara et al., in~\cite{Ferrara2020characterizing}, found how a relatively small number of automated accounts managed to create traffic spikes on the election debate comparable to human users, the latter counting for several orders of magnitude greater. 

Our work focuses precisely on the online Twitter debate in the week before November 4, 2020 and, like other studies, examines the disinformation flow and infiltration of bots into the debate. Unlike other work, however, ours points the binoculars at two specific aspects of the U.S. presidential elections, namely the {\it winner take all} system and the existence of {\it swing and safe states}. In fact, the recent literature, in comparing differences and similarities in the online political debates among different countries, highlights how the different election systems induce different structural properties of online social networks~\cite{Bright2018,Urman2020,VanVliet2021,Praet2021}. 



The term `swing' refers to those states in which one cannot be sure of a landslide victory for Republicans or Democrats since there is not a neat historical orientation of the electorate.
In contrast to swing, a state is called `safe' if its citizens, in recent history, have always elected representatives from the same political party.
In all U.S. states, excluding Maine and Nebraska, the voting methodology is known by the name 
`winner-take-all'. Each state has a  number of presidential electors (dependent, among other things, on the number of residents in the state). 
After a popular election has been held, 
each state converts the popular votes of its citizens into its presidential electors. The `winner-take-all' system implies that whatever the number of votes for the Republican candidate and the Democratic candidate, whoever has the greater number earns the presidential electors of that state.

One of the main criticisms of this system is that it may lead  presidential candidates to focus their campaigns on a few swing states, since they are key states to achieve victory~\cite{Edwards2011why}. In fact, some battleground states have traditionally been the subject of more intense electoral campaigns, let the reader think, e.g., to Florida,  
traditionally a swing state and with a  large population,  thus allocating many presidential electors (29). 


Translating that critique to the Twittersphere, in this paper we ask and answer the following question: is it possible that, in the run-up to the 2020 U.S. presidential election, the Twitter traffic related to the election debate reflects the election system, and, in particular, the division in swing and safe states?


We refer specifically to disinformation flows:
\begin{itemize}
\item  Is the flow of tweets containing links to dubious or non-trustworthy news  different if it involves swing states or safe states? 
\item Does the presence of automated accounts in the pre-election online political debate differ, considering the traffic talking about swing and safe states?
\end{itemize}

To conduct our analysis, we collected Twitter data based on keywords, one of which was always the name of a U.S. state (swing or safe). After filtering out useless data from the entire dataset, we extracted (i) the reputability level of the publishers whose news bounced in the tweets and (ii) the boticity level of the users.





\subsection{Contributions} Our main contributions are:
\begin{itemize}
    \item We provide a fine-grained characterization of the Twitter traffic about the 2020 U.S. presidential elections, in the week leading up to election day, adopting a multidisciplinary approach including complex network analysis, 
    artificial intelligence, 
    and human-based annotation.
    \item To the best of our knowledge, this is the first paper that addresses the problem of understanding whether, in the Twittersphere, the U.S. 2020 pre-election debate was polarized more in tweets about  swing states rather than safe states.
    \item There is evidence of an alignment between the real electoral mechanism -which often favors more intense campaigning in certain locations- and the online electoral debate. 
    \item 2020 election-related traffic focuses, for the vast majority, on tweets about swing states; swing state-related debate sees a higher concentration of links to non-trustworthy news sites. 
   The majority of disinformation content associated with swing states 
is posted and retweeted by automated accounts. 
\end{itemize}




\subsection{Main Results:} 
\begin{itemize}
    
   

    \item Tweets associated with swing states account for about 88\% of the whole traffic.  
    
    
    \item Tweets associated to safe states have a higher concentration of URLs pointing to news with trustworthy publishers. Tweets associated to swing states  have a higher concentration of URLs pointing to news with no trustworthy publishers.

    \item Of the total number of tweets associated with swing states and containing  no trustworthy URLs, 74\% of these are posted or retweeted by accounts classified as bots.

\end{itemize}

\subsection{Originality}
 This work is not the first nor will it be the last to deal with the impact that real events have on virtual events, and vice versa. An in-depth study concerning disinformation flows on social networks in relation to major political events will be presented later in this article. Noteworthy is the work by Howard et al. in~\cite{howard2018social}, which examines tweets from authors who left some evidence of their physical location in the period leading up to the 2016 U.S. presidential election. The result of the analysis shows a high concentration of polarized news in swing states with many presidential electors. 
Aside from the different years (2016 {it vs 2020}), the substantial difference of our study compared to~\cite{howard2018social} is the filtering procedure we apply on our dataset, which employs statistical methods for the analysis of complex networks (suited for the study of social networks interactions) (see Section \ref{sec:rw} and \ref{sssec:DisCo} for further details).  

Using this filtering procedure allows us to bring out political discussions from the data so that we can discard all the rest.
The main idea lies in labeling `standard' users starting from labels assigned to groups of `similar' verified accounts i.e., accounts whose owners are certified by the platform itself. One group of `similar' verified users may be affiliated with a political orientation or not (this check is done manually for each group of verified). So in principle, we will discard (as noise) tweets posted by all `standard' users not affiliated with any political group of verified users. Instead, we will focus our analysis only on tweets posted by `standard' and verified users affiliated with political groups. The importance of focusing on political communities comes from the fact that we are analyzing who is  tweeting about swing and safe states, and what they are tweeting about, but we want to make sure that users are interested in the political narrative, and not, say, in sporting events. The complete filtering procedure will be explained in Section \ref{sssec:DisCo}.

\section{Related Work}
\label{sec:rw}


\paragraph{Disinformation flows in online political debates} 
In the introduction, we highlighted analyses on detecting online disinformation flows in the periods leading up to  the 2016 and 2020 U.S. presidential elections~\cite{Georgacopoulos2020how,Bovet2019influence, shao2018anatomy,Ferrara2020characterizing}.
%
%
Regarding the 2016 elections, the work in~\cite{budak2019happened} stands out from others in analyzing the disinformation flow around candidate Clinton (many works are focused on candidate Trump instead) and highlights how the prevalence of fake news on Twitter increased as the election approached and that the content of such news targeted Hillary Clinton contextually to periods when her online popularity was increasing. 
Luceri et al., in~\cite{Luceri19red}, focus on the 2018 U.S.  midterms and examine the behavior of social bots on Twitter in the period leading up to the event. Social bots are categorized according to their political leanings: conservative bots tend to converse primarily with their human counterparts, while liberal bots are more `democratic', addressing a broader audience.


Online disinformation does not only touch presidential elections, and it does not only touch the United States. Pierri et al., in~\cite{pierri2020investigating}, analyze the online debate on Twitter in the five months leading up to the 2019 European Parliament elections.  In particular, the work looked at well-known Italian disinformation outlets, responsible for the majority of the disinformation circulating on Twitter, with an audience strongly and explicitly related to the Italian far-right political environment.

The work in~\cite{caldarelli2020role}, although not focused on an election event, examines the debate on immigration in Italy, and brings out online communities with specific political identities, reflecting the composition of the Italian parliament at the time of data collection. The work discovers
teams of bots following the most influential accounts in the conservative community and amplifying their messages via retweeting. 

 Despite being a mostly scientific subject, the COVID-19 discussion shows a clear division in what results to be different political groups: Caldarelli et al., in~\cite{Caldarelli2021}, consider the Twitter debate around COVID during the months of the Italian lockdown (March-April 2020) and analyze the news linked in tweets: the vast majority of news from notoriously biased publishers is retweeted online by communities akin to center-right and right-wing political groups. 
 Recent work by Mattei et al.~\cite{mattei2022bowtie} reveals that the retweet networks -where users interact about politics and society- show statistically significant bow-tie structures~\cite{broder2000graph}. When the network is affected by disinformation, the flux to the OUT sector, i.e. the most crowded one in the bow-tie decomposition,  is particularly consistent and display a high frequency of non reputable pieces of news.

 As introduced at the beginning of the article, Howard et al, in \cite{howard2018social} focused on analyzing tweets about swing and safe states posted during  the 2016 U.S. pre-election period. Results show a high concentration of polarized news in swing states with many presidential electors. This work represents a valuable predecessor to our current one. However, we would like to point out that the election we focus on is different, that our work also focuses on social bots, and that, most importantly, we filter out noise from the complete dataset by applying a procedure based on complex network analysis.
\paragraph{Social Bots} 
Research in social bot detection   has been going on for about 12 years. The first work appeared around 2010~\cite{YardiRSB10,mustarafi10}. In the first years of bots hunting ($\sim$2010-2014), researchers mainly focused on supervised machine learning and on the analysis of the single account: `classifiers were separately applied to each account of the group', to which they assigned a bot or not label~\cite{cresci2020decade}.   
Approximately from 2014 to 2019, instead, a number of research teams, independently,
proposed new approaches
for detecting coordinated behavior
of automated malicious accounts, see, e.g.,~\cite{IntSys2015, yu2015}. Thus, researchers no longer took the individual account and classified it as bot or not. Instead, they considered a group of accounts, and their common characteristics, like  anomalies in synchronicity and normality~\cite{Giatsoglou2015,jiang2016}, detection of loosely synchronized actions~\cite{Cao:2014}, or distance between distributions of reputation scores~\cite{viswanath2015}. 

We have seen a convergence of the two techniques (i.e., finding a general-purpose bot detection {\it vs} developing specialized ones for the characteristics of certain bots) in the last 3/4 years. 
A recent work that classifies the single accounts is 
 in~\cite{DBLP:conf/cikm/Sayyadiharikandeh20},  where specialized supervised models are built for various classes of bots. The models are aggregated into an ensemble and their outputs are combined through a voting scheme. 
Yang et al. ~\cite{DBLP:conf/aaai/YangVHM20} handle the full stream of public tweets on Twitter in real time. Key recipe of the analysis is to use account features that have a very low cost in terms of the data needed to compute them. In particular, the authors use features related only to the account profile. The same category of features has been recently tested in~\cite{DENICOLA2021102685} to spot a different kind of bots.

Recent literature has also seen proposals based on coordination as a team.
Hui et al.~\cite{DBLP:conf/icwsm/HuiYTM20} leverage hashtags, links,  phrases, and trending media to let coordinated campaigns emerge from the crowd.
 Other techniques are based, for example, on similarity between sequences of actions~\cite{cresci2018social}, and interactions between accounts~\cite{sharma2020identifying}. 








The main purpose of this paper is not to define a new technique for bot detection, nor is it to detect a particular type of social bot. Therefore, to classify accounts as bots or not, we prefer to adopt Botometer, one of the most well-known tools in the literature for bot unveiling. 
Botometer~\cite{Varol2017,FerraraArming2019} 
is based on a supervised machine learning approach employing Random Forest classifiers~\cite{Breiman2001}. Here, we will rely on Botometer v4, the new version of the bot detector, which has been recently shown to perform well for
detecting both single-acting bots and coordinated campaigns~\cite{DBLP:conf/cikm/Sayyadiharikandeh20}. 

\paragraph{Statistical methods for the analysis of online social networks}
The recent literature regarding online social networks has progressively implemented more techniques based on network science, with the aim of distinguishing  non-trivial signals of social interactions from random noise. In particular, the implementation of entropy-based null-models (see the review by Cimini et al~\cite{Cimini2018a}) has opened up a variety of applications, providing a general and unbiased benchmark for the analysis of complex networks.
The main idea is to create a maximally random benchmark (i.e., maximising the Shannon entropy associated with the system under analysis) that preserves some (topological) property of the original system. In this sense, with the aim of detecting non-trivial behaviours, maximum-entropy null-models represent a tool that, at the same time, is general and tailored on the observed network. 


Here, we provide the sketch of the definition of the entropy-based null models for complex network analysis and all references for further information.



The aim of the entropy-based null-models is to define a benchmark for the analysis of a real network $G^*$ that 
is maximally random, but for a set of topological constraints $\vec{C}$ observed on $G^*$.
Thus, we define an \emph{ensemble} of graphs $\mathcal{G}$, i.e., the set of all possible graph configurations, from the empty graph to the fully connected one, all having the same number of nodes as in the real network. Then, we can assign a probability to every representative of the ensemble by maximising the relative Shannon entropy, i.e.,
\begin{equation*}
    S=-\sum_{G\in\mathcal{G}}P(G)\ln P(G),
\end{equation*}
under the constraint that the average over the ensemble of the vector $\vec{C}$ is exactly the value observed in the real network $G^*$, i.e., $\langle \vec{C}\rangle_\mathcal{G}=\vec{C}(G^*)$.
The result of this procedure returns in an Exponential Random Graph, i.e. $P(G)\sim e^{-\vec{C}(G)\cdot\vec{\theta}}$, where $\vec{\theta}$ are the Lagrangian multipliers associated to the constrained maximisation~\cite{Jaynes1957,park2004statistical}. The maximisation of the likelihood, i.e., the probability of observing the real system, is then implemented to find the numerical values of $\vec{\theta}$~\cite{Garlaschelli2008,Squartini2011a}.

Recently, a fast and efficient Python module able to solve many of the entropy-based null-models present in the literature was released and is available at \href{https://pypi.org/project/NEMtropy/}{\color{blue}\underline{Pypi}}.  

The importance of using a properly defined unbiased benchmark for the analysis of the spread of online disinformation was stressed in a recent work by De Clerck et al.~\cite{DeClerck2022b}: the authors  show how different entropy-based null-models can highlight different features of the various disinformation campaigns. In this paper, we will consider the entropy-based null-model known as Bipartite Configuration Model
(BiCM~\cite{Cimini2018a,Saracco2015}) as a benchmark to maintain only verified Twitter accounts that have statistically significant interactions with unverified ones.
In Section~\ref{sssec:DisCo} we describe the use of this model as a component of our filtering procedure.

\section{Results}

\subsection{Dataset}\label{sec:Dataset}

Using the Streaming Twitter API, we collected around 5.3M tweets in the week immediately preceding the elections (27 October-3 November 2020). To guide the data collection, we chose keywords 
combining the name of four swing and four safe states (see Table~\ref{tab:dataset_by_states}) with the candidates (i.e., Trump and Biden).

\begin{table}[ht!]
\caption{Keywords which drove the data collection phase  
\label{tab:keywords}}

\begin{tabular}{l}
Keywords\\
\hline
\hline
arizona biden\\
arizona trump\\
florida biden\\
florida trump\\
michigan biden\\
michigan trump\\
pennsylvania biden\\
pennsylvania trump\\
new jersey biden\\
new jersey trump\\
indiana biden\\
indiana trump\\
washington biden\\
washington trump\\
louisiana biden\\
louisiana trump\\
\hline
\end{tabular}
\end{table}


States have been chosen based on measures and indications provided in reports by experienced political analysts in the months leading up to the 2020 elections\footnote{\url{https://www.cookpolitical.com/sites/default/files/2020-03/EC\%20030920.4.pdf}}. We opted for a balanced list of states, i.e., four safe and four swing states.
As safe states, we selected two pairs, balanced in terms of political orientation and presidential electors. From the solid Democrats, we took Washington and New Jersey and from the solid Republicans, Indiana and Louisiana. This results in 26 electoral votes for the democratic candidate and 19 votes for the republican one. For the selection of swing states, we took the three most important states from the point of view of presidential electors: Florida (29 votes), Pennsylvania (20 votes) and Michigan (16 votes); we further added Arizona (11 votes) because it aroused particular interest in the election debates\footnote{\url{https://fivethirtyeight.com/features/how-arizona-became-a-swing-state/}}$^,$\footnote{\url{https://www.washingtonpost.com/politics/2022/09/16/senate-control-midterm-elections-2022/}}.


The data were further processed in order to (i)
maintain only accounts that, in the retweet network, have successfully passed the filtering procedure (filtering out useless data from the entire set)
,  (ii) enable link domain classification through NewsGuard, and (iii)  find a mapping between each tweet and the kind of state (i.e., swing or safe). 

As anticipated, the dataset filtering procedure is described in Section~\ref{sssec:DisCo}. From here on, we shall call the `validated dataset' the product of the filtering procedure (to distinguish it from the initial dataset).
For both the verified and unverified accounts that pass the filtering procedure, we also collect the bot scores via BotometerLite.

To enable the classification of URLs, we rely on NewsGuard\footnote{\url{https://www.newsguardtech.com/}}, which provides a set of \{\emph{domain\_name}, \emph{tag}\} pairs (tags are in Table~\ref{table:domains-tags}). It is therefore necessary to translate all the short-form URLs contained in the text of the tweets, so that we can have the domain names in clear. 
We clarify that a domain, for us, corresponds to the so-called `second-level domain' name\footnote{\url{https://en.wikipedia.org/wiki/Domain_name}}, i.e., the name directly to the left of .com, .net, and any other top-level domains. For instance,  \url{nytimes.com} and \url{latimes.com} are considered as domains in the present manuscript.
    
We employ a keyword-based approach to find the association between each tweet and the state type (i.e., swing or safe). In practice, from the text of each tweet -or retweet-, we first check for the presence of at least one state name among the chosen ones (Arizona, Florida, etc.) and then we discard all the tweets in which more than one state appear. Each tweet in the resulting dataset, thus, contains only one state name, which can be either  swing or safe. In addition, we consider English tweets only. The resulting dataset consists of $\sim$3.3M tweets and $\sim$398k URLs (see Table~\ref{tab:dataset_by_states}).

\begin{table}
\caption{Twitter's statistics by state. The asterisk `$\ast$' indicates swing states.}
\begin{tabular}{lcc}
\toprule
State &  No. Tweets &  No. URL \\
\midrule

Arizona$\ast$ &      224046 &    34637 \\
Florida$\ast$  &      744006 &    85373 \\
Michigan$\ast$ &      734600 &    87529 \\
Pennsylvania$\ast$ &     1209083 &   145067 \\

New Jersey &       38007 &     8114 \\
Indiana &       17185 &      988 \\
Washington &      342104 &    36254 \\
Louisiana &        6886 &      633 \\
\bottomrule
Total & 3315917 & 398595 \\
\bottomrule
\end{tabular}
\label{tab:dataset_by_states}
\end{table}

\subsection{Bipartite Configuration Model, Validated Projection and Community Detection}\label{sssec:DisCo}


Here, we describe how we filter accounts in our dataset using the validation procedure known in the literature as Bipartite Configuration Model BiCM~\cite{Cimini2018a,Saracco2015}.
As anticipated, our aim is to bring out political communities, leveraging the knowledge of the  political affiliation of verified users.

The first observation is that most of the online debate is led by verified users, i.e., accounts whose owners are certified by the platform itself~\cite{Becatti2019d,Caldarelli2021,caldarelli2020role}. It is possible, therefore, to leverage this information to obtain proper communities of `similar' verified users: the intuition is that verified users with similar opinions in an online debate should have the same audience of `standard' users. 

Therefore we represent the retweet interactions between verified and unverified users as a bipartite network, i.e., networks in which nodes are divided in two sets, $\top$ and $\bot$ -called \emph{layers}- and connections are allowed only between layers; verified and unverified users are then represented by the two layers.



We then project the bipartite network on the layer of verified users. Nevertheless, the projection only does not tell us so much: In fact, the common retweeters of two verified users could be many due to popularity of the latter or because the retweeters are retweeting many verified users.  We, therefore, need a benchmark that is maximally random and able to discount the effect of these two ingredients, which, in terms of the bipartite network defined above, are translated into the degree sequence of both layers. The entropy-based null-model for bipartite network discounting the information of the degree sequence is known as \emph{Bipartite Configuration Model} (BiCM,~\cite{Saracco2015}).

Using the BiCM as a benchmark, it is possible to validate the projection of the bipartite network on one of its layers: the co-occurrences observed in the real system are compared with the related BiCM distributions and, if they are statistically significant, they are validated~\cite{Saracco2017}. Therefore, the result of the validation procedure is a monopartite undirected unweighted network of verified users, in which two nodes are connected if the number of common retweeters is statistically significant, i.e., {\it it cannot be explained simply by the bipartite degree sequence.}

We subsequently run the Louvain community detection algorithm~\cite{Blondel2008} on the validated network of verified to obtain the main communities. Each of these communities was manually labeled based on the characteristics of the  verified users inside.



Then, to include also unverified users, the so-obtained labels are propagated on the retweet network using the Raghavanan et al. algorithm~\cite{Raghavan2007b}, in order to provide all users a 
community label. Several works, like~\cite{Becatti2019d,caldarelli2020role,Radicioni2021a,DeClerck2022a,DeClerck2022b} show that the procedure above is particularly effective in capturing the structure of Twitter online debate.

After the label propagation, the biggest community is mostly composed by Republican supporters. The second most populated community is a mixed one, including Republicans, Democrats and some journals and journalists with various political orientations. 

\begin{figure}[h]
  \centering
  \includegraphics[width=\linewidth]{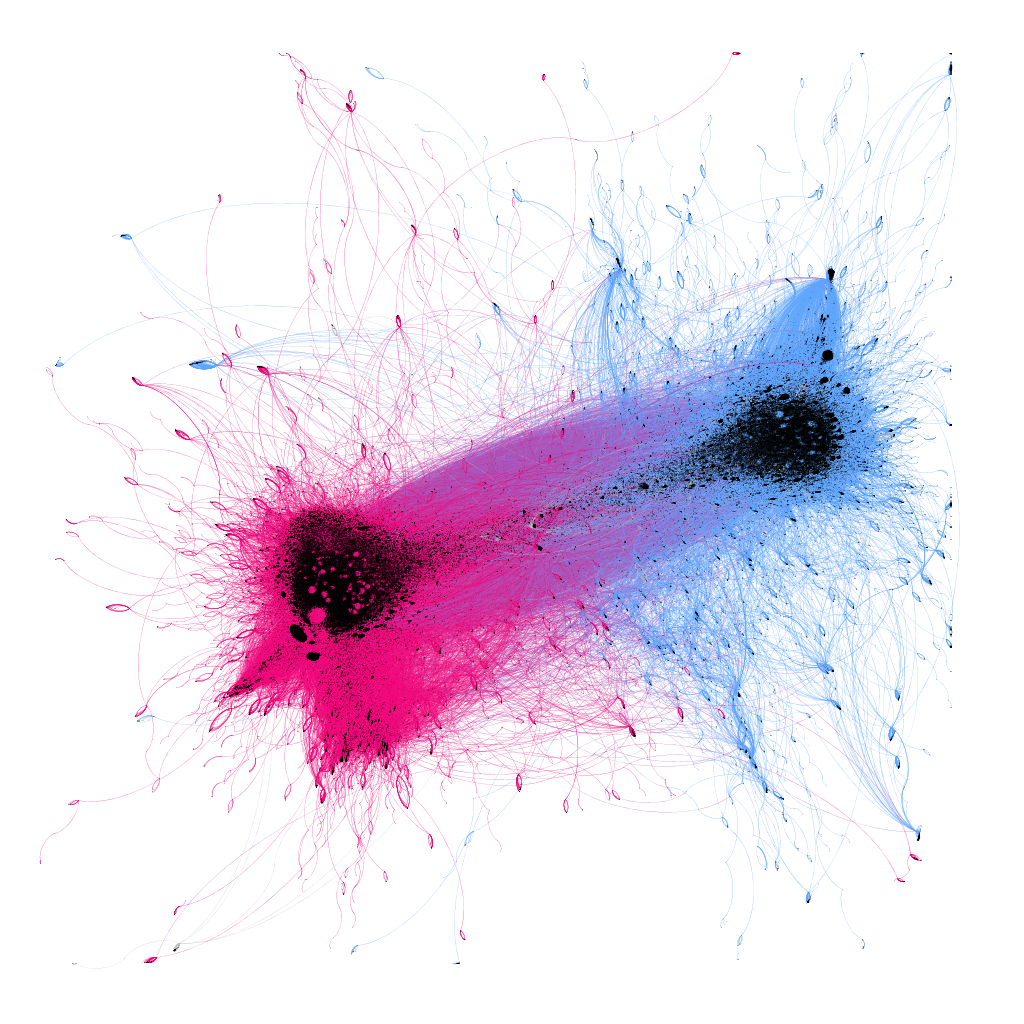}
  \caption{Retweet Network after label propagation (547k nodes, 1.8M edges). The two main communities that emerge are the red one, characterised by Republican supporters and the blue one, a mix between Republicans, Democrats, and journalists of various affiliations.
  \label{fig:retweet_network}}
  \Description{}
\end{figure}

Figure~\ref{fig:retweet_network} shows the retweet network in terms of these two main communities.

\subsection{Reputation of news domains}
All the domains in our validated Twitter dataset have been tagged according to their degree of credibility and transparency,  as indicated by the browser extension and mobile app NewsGuard.
%
For evaluating the credibility level, the Newsguard metrics consider, e.g., whether the news source regularly publishes false news, does not distinguish between facts and opinions, does not correct a wrongly reported news. For transparency, instead, the toolkit takes into account, e.g., whether owners, founders or authors of the news source are publicly known, and whether advertisements are easily recognizable\footnote{Details on the procedure for the  evaluation  are available at: \url{https://www.newsguardtech.com/ratings/rating-process-criteria/}.}

\begin{table}[ht!]
\caption{Tags for domain reputation labeling. Tags are inherited from NewsGuard, the UNC tag indicates that NewsGuard has not yet tagged that domain. 
\label{table:domains-tags}}
\centering
\begin{tabular}{c|l}
label & \text{description}\\
\hline
\hline
T & Trustworthy news domain\\
N & Non-trustworthy news domain\\
P & Platform (e.g., reddit.com, twitter.com)\\
S & Satire\\
UNC & unclassified\\
\hline
\end{tabular}

\smallskip

\end{table}

Table~\ref{table:domains-tags} shows the tags associated to the domains provided by NewsGuard. We are interested in quantifying reputation of news domains publishing during the period of interest. 
Thus, we do not consider those sources corresponding to platforms (tag P). Also, we will not consider satiric news (tag S). 
Tags T and N in Table~\ref{table:domains-tags} are used only for news sites, be them newspapers, magazines, TV or radio social channels, and they stand for Trustworthy and Non-trustworthy, respectively.

\begin{figure}[h]
  \centering
  \includegraphics[width=\linewidth]{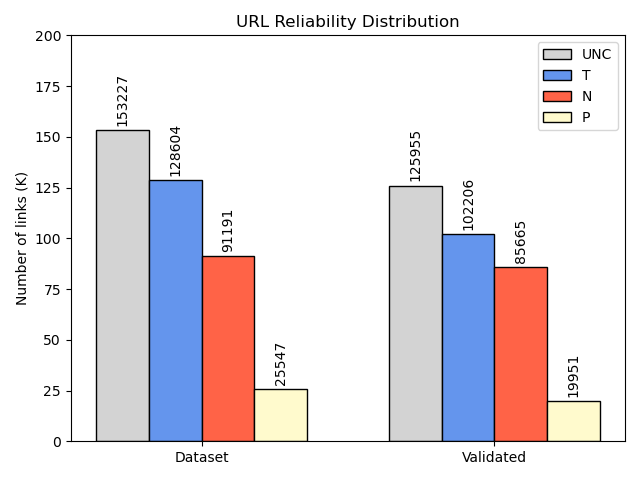}
  \caption{Classification of links: trustworthy news publisher (T); non-trustworthy news publisher (N); not a news site, e.g. platforms as amazon.com (P); link corresponding to a domain not covered by NewsGuard (UNC).\label{fig:url_reliability_distribution}}
\end{figure}



Figure~\ref{fig:url_reliability_distribution} shows the distribution of URLs found in tweets in the complete and validated dataset regarding the tags that NewsGuard assigned to the related domains. We can observe that the validation procedure discards $\sim17\%$ of the tweets from the complete dataset, which translates into approximately 64k tweets. Hence, most links are disseminated within the political communities that emerge from data.

\subsection{Bot detection}
\label{sec:botdetection}
The accounts in our dataset were examined with the bot detector Botometer~\cite{Varol2017,FerraraArming2019,DBLP:conf/cikm/Sayyadiharikandeh20}. The tool is based on a supervised machine learning approach employing Random Forest classifiers~\cite{Breiman2001}. 

In particular, we adopted Botomoter v4 premium in the lite version BotometerLite\footnote{\url{https://cnets.indiana.edu/blog/2020/09/01/botometer-v4/}},  which does not interface with Twitter, but simply takes the tweet, retrieves the author, and does the necessary follow-up analysis. This light version only needs the information in the user profile to perform bot detection and hence can also process historical data published by accounts that are no longer active. Each request to BotometerLite can process a maximum of 100 users, with the limit of 200 requests per day, leading to a maximum of 20k account checks per day.

%
The immediate output of Botometer is the bot score $S$ ranging over \{0, \ldots 1\}, which however does not represent the probability that the considered account is a bot. The value has to be compared with other scores within a group of accounts, to come up with a plausible ranking.


\subsection{Reputation of news domains in tweets associated to swing and safe states}
\label{sec:reputationState}

\begin{table}[]
\caption{Statistics regarding number of accounts, tweets, and type of URLs per state type in \sc{Validated dataset}.
\label{tab:misinformation_in_swing_and_safe}}
\begin{tabular}{lccccc}
\toprule
States &  No. Users &  No. Tweets &  No. URL &  T &  N \\
\midrule

Swing &     636125 &     \textbf{2911735} &   352606 & \textbf{29.84} & \textbf{23.47} \\ 
Safe &     214329 &      \textbf{404182} &    45989 & \textbf{50.87} & \textbf{18.33} \\ 

 
\bottomrule
\end{tabular}
\end{table}
Here, we analyze the flow of disinformation in tweets associated with swing or safe states. We recall that a tweet is associated with a state if the name of the state is in the tweet text. Each tweet in our dataset contains only one state name, by construction.



Table~\ref{tab:misinformation_in_swing_and_safe} gives statistics regarding number of accounts, tweets, and URLs w.r.t. the state associated to the tweets. 
We observe that the vast majority of traffic is associated with tweets about swing states (about 88\% of the total, see row Dataset, column No. Tweets). Considering links pointing to non-trustworthy news sites (N), the  concentration for swing states - 23.47\% - is higher than that for safe ones - 18.33\%. The  concentration of trustworthy links (T) is higher for safe states - 50.87\% {\it vs} 29.84\% for swings.





\subsection{Social bots}



In this section, we investigate the relationship between disinformation flow and the nature of the accounts in our dataset.  We calculate the bot scores of the accounts using BotomerLite (details in Section~\ref{sec:botdetection}). The bot score provides a measure of how much an account has bot-like features, on a scale between 0 and 1 -the closer the score to 1, the more likely the account is bot.

To determine which accounts are bots, we choose the conservative approach used in~\cite{Ferrara2020characterizing}: we classify as bots those accounts `that sit at the top end of the bot score distribution'. The advantage is twofold: on the one hand, we avoid misclassifying accounts with a borderline score; on the other hand, we focus on accounts that exhibit clear bot characteristics. 
Practically, we tag each account in the validated dataset via BotometerLite, we sort them from the lowest to the highest bot score, and isolate those with bot scores in the first and last decile. 
In the first decile we have genuine accounts, in the last decile we have bot accounts. The first decile includes accounts with bot score in [0, 0.04]; the last decile include accounts with bot score in [0.45, 1].

We take the tweets of each of the genuine accounts and the bots, with the aim of investigating responsibility of the non-trustworthy traffic attributable? Obviously, we are cutting many accounts from our validated dataset, since we do not consider those with bot scores in \{0.04, 0.45\}. 
Table~\ref{tab:human_bot_user_classification} shows some statistics for only the accounts that we have classified. 
Out of the total number of classified accounts, bots account for 47.19\%. 

\begin{table}
\caption{Genuine and bot accounts in the validated dataset \label{tab:human_bot_user_classification}}
\begin{tabular}{lccc}
\toprule
        Label &  No. Users &  No. Tweets &  No. URL \\
\midrule
        \multicolumn{4}{l}{\sc{Validated dataset}} \\ 
        human &      57797 &      228378 &    25422 \\
        bot &      51648 &      409449 &    53017 \\
\bottomrule
\end{tabular}

\end{table}


In terms of posts, bots appear to be more active than genuine accounts (about twice as active) in both posting tweets and tweets with URLs. 
 Out of the total traffic from classified accounts, 
 bots produce 64.19\% of the traffic. 

We concentrate on the role played by bots in spreading links to low/non-trustworthy news stories. Table~\ref{tab:misinformation_for_bot_and_states} shows the percentages of (i) all, (ii) trustworthy (T) and (iii) non-trustworthy (N) URLs shared by users classified as bot or genuine. The table also considers the belonging to 
a state category (swing or safe). 



    Focusing on the \emph{Swing \& Safe} column in Table~\ref{tab:misinformation_for_bot_and_states}, we see that about the 73\% of the non-trustworthy (N) traffic is generated by bots, 
    while, on the other hand, they are responsible for about $\sim63\%$  of tweet displaying trustworthy URLs.  

\begin{table*}[]
    \caption{Percentages of links shared, per reputability and per state type. \label{tab:misinformation_for_bot_and_states}}

\begin{tabular}{lcccccccccccc}
\toprule
    &&&&& \multicolumn{2}{c}{\emph{\textbf{Swing \& Safe}}} && \multicolumn{2}{c}{\emph{\textbf{Swing}}} && \multicolumn{2}{c}{\emph{\textbf{Safe}}}\\
    \cmidrule{6-7}
    \cmidrule{9-10}
    \cmidrule{12-13}
    
    Link type &  No. URL &  swing &  safe &&  bot &  human &&  bot &  human &&  bot &  human \\
    \midrule
    {\sc All Links} & 78439 &  89.92 & 10.08 &&      \textbf{67.59} &        32.41 &&   \textbf{67.87} &     32.13 &&   \textbf{65.06} &     34.94 \\
    \midrule

    {\sc Trustworthy Links (T)} & 23036 &  83.07 & 16.93 &&      \textbf{62.90} &        37.10 &&   62.69 &     37.31 &&   63.96 &     36.04 \\
    \midrule

    {\sc Non-trustworthy Links (N)}& 20627 &  \textbf{91.53} &  \textbf{8.47} & &     \textbf{73.69} &        26.31 &&   \textbf{74.15} &     25.85 &&   \textbf{68.75} &     31.25 \\
\bottomrule
\end{tabular}
   
\end{table*}

If we focus on non-trustworthy links only, of the 91\% of the total in swing states, more than 74\% are posted or retweeted by bots. Furthermore, although non-trustworthy links associated with safe states are only a small part of the total (8.47\%), the vast majority of this traffic comes from bots accounts (68.75\%).

\section{Discussion}
The analysis of disinformation in online social networks during election campaigns features several contributions, like~\cite{Becatti2019d, Bovet2019influence, budak2019happened, Ferrara2020characterizing, Georgacopoulos2020how,luceri2019evolution, mattei2022bowtie} to cite a few. Nevertheless, the spreading of non reputable content has been rarely compared to the peculiarities of a specific election system; most of the existing studies on disinformation focus on a single country. However, the election procedure seems to have a role in the way the online discussions evolve: the few results available so far~\cite{Bright2018,Urman2020,VanVliet2021,Praet2021, howard2018social} indicate that there are indeed some differences in the way accounts organise in online debates, i.e, according to more divisive or cohesive structures, in countries with majoritarian, proportional or plurality election systems.

In the present work, we still consider a single country, but we focus on (i) a characteristic of its presidential election system, i.e., the presence of swing and safe states and (ii) if and to which extent  this feature is reflected in the online disinformation spreading.

More in detail, each U.S. state  has a  number of presidential electors and, after the popular elections at state level, the faction that obtains the greatest number of votes, earns all of them, independently on the margin of the final result. Then, \emph{safe} states are those in which the electorate has a traditional orientation and the result of the election can be easily predicted, while \emph{swing} states are the ones to compete for in order to obtain the presidential election.

Therefore, we focused our analysis on the 2020 U.S. presidential elections, and we consider the Twitter debate associated with 8 states, 4 of them being safe ones (New Jersey, Indiana, Washington and Louisiana), the other 4 being swing ones (Arizona, Florida, Michigan and Pennsylvania). Then, we selected tweets displaying in their text the name of the presidential candidates (either Biden or Trump) and the name of one of the selected states.

Our first result is that 88\% tweets in our dataset is related to swing states. If we consider the population of the various states as a proxy for the estimate of the relative amount of traffic of the swing state, we will expect a percentage of 66\%: the mismatch witnesses a greater attention on the election campaigns on these states.

Secondly, from Table~\ref{tab:misinformation_in_swing_and_safe we observe that} the frequency of non-trustworthy URLs shared in the political debate of swing states (23.47\%) is greater than the analogous of safe states (18.33\%). Symmetrically, the frequency of trustworthy URLs is higher in safe states (50.87\%) than swing ones (29.84\%). 
In this sense, not only the debate, but also the diffusion of disinformation, is more intense in swing states, due to their importance for the election outcome. Summarising, we have that both the total flux of messages and the frequency of non reputable URLs are higher in swing states.


Thirdly, we investigate the contribution of automated accounts in the spreading of disinformation. Let the reader consider Table~\ref{tab:misinformation_for_bot_and_states}: bots appear to be more active than genuine accounts in posting tweets, both in swing and in safe states, with comparable percentages, i.e. $\sim$67\% vs. $\sim$65\%, respectively in swing and safe states. Regarding the non-trustworthy links shared in swing states, more than 74\% are posted or retweeted by bots.

Our analyses were carried out by applying filtering to the initial dataset. We used techniques based on Information Theory and Statistical Mechanics of complex networks (see Section~\ref{sssec:DisCo}) to bring out political communities. In particular, we focused on the bipartite network that represents the retweet interaction between verified and unverified users. Using the BiCM as a benchmark we validate the projection of the bipartite network on the layer of verified users: we put a link between two of them if the number of common unverified retweeters is statistically significant. We then ran a community detection algorithm over the so-obtained network of verified users; we extended the communities to unverified Twitter users too, by exploiting the knowledge about verified ones and a label propagation procedure. With our validation method, we ensure that we consider interactions that cannot be explained by the degree sequence of the users. We emphasize the application of this filtering, which differentiates us from other work, as that in~\cite{howard2018social}, which analyzes disinformation flows in swing and safe states in 2016, but without applying entropy-based null models.

Summarising, our hypothesis that the diffusion of disinformation is more intense in swing states is confirmed by data: due to their relevance for the outcome of the election, swing states both attract more tweets and, in percentage, are more exposed to disinformation campaigns than safe states. The relative impact of disinformation and the stronger flux of messages result in a particularly worrisome flux of disinformation messages. 

\paragraph{Limitations and future work}
While our results are neat, there are still some limitations that call for further studies. First, we investigate only a limited amount of U.S. swing and safe states. Then, we just analyzed the 2020 U.S. presidential election, while a comparison with the 2012 and 2016 elections could confirm our conclusions or limit them to the 2020 competition only.\\
Moreover, by adopting a keyword-based data collection method, we do not know \emph{a priori} the exact content of the collected tweets (although, since there is both the state name and the candidate name, the tweet plausibly refers to the election and that state).

Further, other plurality election systems have similarities with U.S., as the U.K. one: it would be interesting to study if analogous diffusion of disinformation on swing electoral constituencies are present. It would also be interesting to consider if the diffusion of disinformation at geographical level is present also in other election systems, featuring, e.g.,  the proportional (as in Germany and Spain), majoritarian (as in France) or mixed modes (as in Italy, South Korea and Japan). 
We argue we have contributed to a finer granularity study concerning the link between electoral systems, online debates, and the presence of online disinformation, and we release the dataset for the benefit of the scientific community.

\begin{acks}

This work was partially supported by project SERICS (PE00000014) under the MUR National Recovery and Resilience Plan funded by the European Union - NextGenerationEU, by the Integrated Activity Project TOFFEe (TOols for Fighting FakEs) \url{https://toffee.imtlucca.it/} and by the IIT-CNR funded Project re-DESIRE (DissEmination of ScIentific REsults 2.0).

\end{acks}

\bibliographystyle{ACM-Reference-Format}

\appendix

\section{Online Resources}
The Twitter datasets used and analysed in the current study are here: \url{https://doi.org/10.7910/DVN/ANBPTC}. The data about the reliability of the various news sources -that support the findings of this study-  comes from NewGguard, but restrictions to their availability apply, since they were used under a NewsGuard license and they are not publicly available. These data could be however available upon reasonable request and with permission of Newsguard.

\end{document}